\documentclass[12pt]{article}

\usepackage[nosort]{cite}
\usepackage{graphicx}
\usepackage{color}
\usepackage{amsmath}
\usepackage{epsfig}
\usepackage{amsfonts}
\usepackage{amssymb}
\usepackage{amsthm}
\usepackage{array}
\usepackage{axodraw}
\usepackage{slashed}

\newcommand{\be}{\begin{equation}}
\newcommand{\ee}{\end{equation}}
\newcommand{\bee}{\begin{equation*}}
\newcommand{\eee}{\end{equation*}}
\newcommand{\bea}{\begin{eqnarray}}
\newcommand{\eea}{\end{eqnarray}}
\newcommand{\bean}{\begin{eqnarray*}}
\newcommand{\eean}{\end{eqnarray*}}

\begin{document}

\setcounter{page}{0}
\thispagestyle{empty}


\begin{center}
{\bf \LARGE { On Neutrino Masses in the MSSM with \\ \vskip0.5cm BRpV}}
\end{center}

\vskip 12pt

\begin{center}
{  Marco A. D\'iaz$^{a}$,  Maximiliano A. Rivera$^{b}$ and Nicol\'as Rojas$^{a}$}
\end{center}

\vskip 20pt

\begin{center}

\centerline{$^{a}${\it Departamento de F\'{\i}sica, Universidad Cat\'olica de Chile,}}
\centerline{{\it Avenida Vicu\~na Mackenna 4860, Santiago, Chile}}
\centerline{$^{b}${\it Departamento de F\'isica, Universidad T\'ecnica Federico Santa Mar\'ia}}
\centerline{{\it Casilla 110-V, Valparaiso, Chile}}

\vskip .3cm
\centerline{\tt mad@susy.fis.puc.cl, maximiliano.rivera@usm.cl, nrojas1@uc.cl}
\end{center}

\vskip 13pt

\begin{abstract}
One loop corrections to the neutrino mass matrix within the MSSM with Bilinear R
Parity Violation are calculated, paying attention to the approach in which an 
effective $3\times 3$ neutrino mass matrix is used. The full mass matrix
is block diagonalized, it is found that second and third order terms 
can be numerically important, and this is analytically understood. Top-stop
loops do not contribute to the effective $3\times 3$ at first order, nevertheless
they contribute at third. An improved $3\times 3$ approach that include
these effects is proposed. A scan over parameter space is made supporting the 
conclusions.

\end{abstract}

\renewcommand{\theequation}{\arabic{section}.\arabic{equation}} 

\section{Introduction}

The  evidence for neutrino oscillation comes from many experiments around the 
world \cite{Fukuda:1994mc,Fukuda:1998mi,Eguchi:2002dm,Ahn:2002up,Sanchez:2003rb,
Ahmad:2001an,Abdurashitov:2002nt,Hampel:1998xg,Ambrosio:2003yz}. The  activity around
neutrino physics has grown due to a more precise determination of neutrino oscillation
 parameters, specially coming from experiments connected with the reactor angle
$\theta_{13}$ \cite{T2K, MINOS, DayaBay, RENO,DChooz}. Global fits \cite{neutrinoData}
using data from the mentioned experiments, allow to extract three mixing angles: two 
large $\theta_{21}$ and $\theta_{23}$, one small $\theta_{13}$, and 
two mass scales $\Delta m^2_{21}$ and $\Delta m^2_{32}$.   
This information,  constitutes an experimental evidence that the 
Standard Model (SM) must be extended.

If neutrinos are massive Majorana particles, lepton number violating
terms  must  be present. In  the Minimal Supersymmetric Standard Model (MSSM)
\cite{Haber:1984rc} with Bilinear  R-Parity Violation  (BRpV) \cite{Farrar:1978xj},  
R-parity is broken via lepton number violation, introducing a bilinear term at 
the superpotential level 
\cite{Barbier:2004ez,Diaz:1997xc,Hirsch:2000ef,Diaz:2003as}.  Therefore, neutrino  masses and mixing 
angles are generated via a low-energy see-saw mechanism, mixing neutrino
flavor-eigenstates and neutralinos.
Although this solution is appealing to explain neutrino masses and
mixing angles, signals for supersymmetry at the LHC have not been seen \cite{Chatrchyan:2011zy}.
Since the majority of the searches are based on supersymmetry with 
bilinear R-parity conserved, there is an open window for it.

In the MSSM with R-Parity violation, one neutrino mass is generated at
tree-level, while  the other two neutrinos remain massless. To 
reconcile theoretical predictions with the
experimental data requires going beyond the tree-level approximation
\cite{Hempfling:1995wj}.
Several authors  have shown the  dependence of the neutrino  masses in
terms of the parameter which bilinearly violate R-parity, and also how
to   determine  these   from   collider  physics   \cite{Porod:2000hv}.
Improvements in  the precision measurement  of the neutrino parameters
\cite{Minakata:2004jt}, as it will be discussed, suggest  to  go beyond
one loop  order  in  the  calculation  of the  neutrino  masses.

The most convenient way to numerically introduce one loop corrections to
neutrino masses in this model is through the $7\times 7$ mass matrix, which
includes 4 neutralinos and 3 neutrinos. If this
mass matrix is block diagonalized, an effective $3\times 3$ neutrino mass
matrix is generated, and it is very convenient when an algebraical understanding
is sought. Nevertheless, the $3\times 3$ approach can miss important 
numerical effects. This motivates a more careful treatment of the block
diagonalization, leading to an improved $3\times 3$ approach.

The paper is organized as follow: In section \ref{sec:BRpV},
introductory remarks about neutrino mass generation in BRpV are provided. 
Section \ref{sec:HOE} shows
how loop corrections are treated in this article. Section \ref{sec:algebra} 
develops algebraic approximations that explain the numerical effects. 
Finally, conclusions about the findings are provided.

\section{Neutrino Masses in Bilinear R-Parity Violation} \label{sec:BRpV}

Models with BRpV include a bilinear term in the superpotential that
violates simultaneously R-Parity and lepton number. The superpotential
has the following form,
\begin{equation}
W=W_{Yuk}+\varepsilon_{ab}\left(-\mu\widehat H_d^a\widehat H_u^b
+\epsilon_i\widehat L_i^a\widehat H_u^b\right),
\label{Superpotential}
\end{equation}
where in $W_{Yuk}$ one has the usual R-Parity conserving (hereafter, RpC) Yukawa terms.
Here the explicit bilinear terms are shown, with $\mu$ the higgsino mass
and $\epsilon_i$ the BRpV mass parameters. In this work trilinear R-Parity 
violating terms are not considered, motivated by models that generate BRpV 
and not TRpV \cite{Mira:2000gg}.
The terms shown in eq.~(\ref{Superpotential}) induce a mixing between neutralinos 
and neutrinos, forming a set of seven
neutral fermions $F^0_i$. The corresponding tree level mass terms can be written 
by a $7\times7$ mass matrix as follows,
\begin{equation}
{\cal M}_N^0=\left[\begin{array}{cc} 
{\mathrm M}_\chi^0 & m^T \\ 
m & 0  \end{array}\right].
\label{X07x7}
\end{equation}
The sub-matrix ${\mathrm M}_\chi^0$ is the usual tree-level neutralino mass matrix of the
MSSM, and $m$ is the BRpV mixing matrix which mix neutralinos and neutrinos. Those are
given by,
\begin{eqnarray}
{\mathrm M}_\chi^0=\left[\begin{array}{cccc}
M_1 & 0 & -\frac{1}{2} g' v_d & \frac{1}{2} g' v_u \\
0 & M_2 & \frac{1}{2} g v_d & -\frac{1}{2} g v_u \\
-\frac{1}{2} g' v_d & \frac{1}{2} g v_d & 0 & -\mu \\
\frac{1}{2} g' v_u & -\frac{1}{2} g v_u & -\mu & 0
\end{array}\right]
&\,\,\,;\,\,\,\,&
m=\left[\begin{array}{cccc}
-\frac{1}{2} g' v_1 & \frac{1}{2} g v_1 & 0 &\epsilon_1 \cr
-\frac{1}{2} g' v_2 & \frac{1}{2} g v_2 & 0 & \epsilon_2 \cr
-\frac{1}{2} g' v_3 & \frac{1}{2} g v_3 & 0 & \epsilon_3
\end{array}\right].
\label{X0massmat}
\end{eqnarray}
The matrix $m$ includes the sneutrino vacuum expectation values $v_i$. These vev`s
appear induced by the $\epsilon_i$ in the superpotential as well as by the corresponding 
soft bilinear terms, not shown in this article (for more details, see \cite{Hirsch:2000ef,Hirsch:2000jt}).
Eq. (\ref{X07x7}) can  be block-diagonalized using the rotation matrix,
\begin{equation}
{\bf R}^0_{bd}=\left[\begin{array}{cc}
1 - \frac{1}{2}\xi^T\xi & \xi^T \\
-\xi & 1 - \frac{1}{2}\xi\xi^T
\end{array}\right],
\label{R0}
\end{equation}
with $\xi=m \, {{\mathrm M}_\chi^0}^{-1}$. In this way, the block-diagonal mass matrix is,
\begin{eqnarray}
{\cal M}_N^{bd,0} &=& \left[\begin{array}{cc} 
{\mathrm M}_\chi^0+
\frac{1}{2}(m^T m \, {{\mathrm M}_\chi^0}^{-1}+{{\mathrm M}_\chi^0}^{-1} m^T m) & 0
\\ 0 &  -m \, {{\mathrm M}_\chi^0}^{-1} \, m^T\end{array}\right]
\nonumber\\
&\equiv& 
\left[\begin{matrix}
{\mathrm M}_\chi^{bd,0} & 0 \cr 0 & {\mathrm M}_\nu^{bd,0}
\end{matrix}\right].
\label{X07x7bd}
\end{eqnarray}
The correction in the neutralino sector is usually ignored, while the correction 
in the neutrino sector is the well known tree-level neutrino effective mass matrix,
\begin{equation}
{\bf M}^{bd,0}_\nu=-m\,{{\mathrm M}_\chi^0}^{-1}\,m^T=
\frac{ M_1 g^2 + M_2 g'^2 }{4\det{{\mathrm M}_\chi^0}}
\left[\begin{array}{cccc}
\Lambda_1^2        & \Lambda_1\Lambda_2 & \Lambda_1\Lambda_3 \cr
\Lambda_2\Lambda_1 & \Lambda_2^2        & \Lambda_2\Lambda_3 \cr
\Lambda_3\Lambda_1 & \Lambda_3\Lambda_2 & \Lambda_3^2
\end{array}\right],
\label{treenumass}
\end{equation}
with $\Lambda_i=\mu v_i+\epsilon_i v_d$. The matrix clearly has only one
eigenvalue different from zero, which is experimentally unacceptable.

It is known that this problem is solved by radiative corrections. Concentrating only
on loops with neutrinos in the external legs, one has for example,
\begin{center}
\vspace{-50pt} \hfill \\
\begin{picture}(200,120)(0,23) 
\ArrowLine(20,50)(80,50)
\Text(50,60)[]{$\nu_j$}
\ArrowArcn(110,50)(30,180,0)
\Text(110,10)[]{$S^0_\ell$}
\DashCArc(110,50)(30,180,0){3}
\Text(110,95)[]{$F^0_k$}
\ArrowLine(140,50)(200,50)
\Text(170,60)[]{$\nu_i$}
\end{picture}
\vspace{30pt} \hfill \\
\end{center}
\vspace{-10pt}
where $F^0_k$ are the mentioned neutral fermions and $S^0_\ell$ are scalars formed by the mixing
between Higgs bosons and sneutrinos \cite{Hirsch:2000ef}. These contributions can be calculated approximately 
in the block-diagonalized basis, obtaining a generalization to the neutrino mass matrix in 
eq.~(\ref{treenumass}), which is customary to write as,
\begin{equation}
\big[ {\bf M}^{bd(1)}_\nu \big]_{ij} = A \, \Lambda_i \Lambda_j +
B \big( \Lambda_i \epsilon_j + \Lambda_j \epsilon_i \big) + C \, \epsilon_i \epsilon_j\,,
\end{equation}
where the parameter $A$ receives tree-level contributions given in eq.~(\ref{treenumass}), while the parameters 
$B$ and $C$ are loop generated. It is also worth mentioning that the parameter $C$ is scale invariant, while 
$B$ is not. 

As mentioned, the neutrino/neutralino tree-level mass matrix is completely diagonalized. 
This is done by applying an extra rotation to the one shown in eq.~(\ref{R0}). This is,
\begin{equation}
{\bf R}^0_{xd}=\left[\begin{array}{cc}
N & 0 \\
0 & N_\nu \end{array}\right].
\label{R0xd}
\end{equation}
The matrix $N_\nu$ diagonalizes the effective tree-level neutrino mass matrix given in 
eq.~(\ref{treenumass}) \cite{Diaz:2003as}, and the $N$ matrix diagonalizes the $4\times4$ neutralino mass matrix.
The net effect is to have,
\begin{equation}
\mathcal{M}_N^{d,0} =
{\bf R}^0_{xd} \, {\bf R}^0_{bd} \, {\cal M}_N^0 \, {\bf R}^{0T}_{bd} \, {\bf R}^{0T}_{xd} =
\left( \begin{array}{rr}
{\mathrm M}_\chi^{d,0} & 0 \\ 0 & {\mathrm M}_\nu^{d,0} \end{array} \right).
\end{equation}
It is at this point that quantum corrections are included,
\begin{eqnarray}
\mathcal{M}_N^1 =\mathcal{M}_N^{d,0} + \Delta \mathcal{M}_N^1 &=& \left( \begin{array}{cc}
                      {\mathrm M}_\chi^{d,0} + \delta M_\chi & \delta m^T \\
                         \delta m & {\mathrm M}_\nu^{d,0} + \delta M_\nu
                    \end{array} \right), \label{eq:BDcorrections}
\end{eqnarray}
where $\delta M_\chi$ are one-loop corrections within the neutralino $4\times4$ 
sub-matrix, $\delta M_\nu$ the one-loop corrections to the $3\times3$ neutrino sub-matrix, and
$\delta m$ refers to the one-loop corrections to the neutralino/neutrino mixing sector.
The above matrix can be block-diagonalized again, obtaining the following result,
\begin{equation}
{\cal M}_N^{bd,1} = \left[
\begin{array}{cc} 
{\mathrm M}_\chi^{bd,1} & 0 \\ 0 & {\mathrm M}_\nu^{bd,1} \end{array} \right],
\end{equation}
where there have been defined,
\begin{equation}
{\mathrm M}_\nu^{bd,1} =
{\mathrm M}_\nu^{d,0}+ \delta M_{\nu} 
- \delta m \,({\mathrm M}^{d,0}_\chi)^{-1}\, \delta m^T
+ \delta m \,({\mathrm M}^{d,0}_\chi)^{-1}\, \delta M_{\tilde{\chi}}\,
({\mathrm M}^{d,0}_\chi)^{-1} \delta m^T \label{OneLoopCorr}
\end{equation}
and
\begin{equation}
{\mathrm M}_\chi^{bd,1} = {\mathrm M}_\chi^{d,0} + \delta M_\chi
\end{equation}
Notice that the last two terms in equation (\ref{OneLoopCorr})
are of second  and third order in our block-diagonalization expansion, and thus they are susceptible 
to be neglected. Nevertheless, since the neutrino masses are several orders of magnitude smaller than 
the neutralino masses, the two terms are numerically important.

%
\section{High Order Effects on Neutrino Masses} \label{sec:HOE}
%

In order to show these effects, one-loop corrected neutrino masses 
in a specific supersymmetric scenario are calculated.
A few of the parameters that define this benchmark are shown in Table \ref{tab1},
\begin{table}
\begin{center}
\begin{tabular*}{0.4\textwidth}{@{\extracolsep{\fill}} c r c }
\hline\hline
Parameter   & Value   & Units \\ \hline
$\tan\beta$ & $16.7$  &  -    \\
$\mu$       & $3171$  & GeV   \\
$M_1$       & $409$   & GeV   \\
$M_2$       & $587$   & GeV   \\
$M_3$       & $5240$  & GeV   \\
$M_Q$       & $4436$  & GeV   \\
$M_U$       & $4037$  & GeV   \\
$M_D$       & $4668$  & GeV   \\
$M_L$       & $1668$  & GeV   \\
$M_R$       & $1964$  & GeV   \\
\hline\hline
\end{tabular*}
\caption{Supersymmetric parameters at the renormalization scale $Q=4233$ GeV. Sfermion mass parameters 
are given for the third generation.
\label{tab1}}
\end{center}
\end{table}
where the given scalar masses correspond to the third generation. In addition, in Table \ref{tab2}
are shown the masses of a few relevant particles.
\begin{table}
\begin{center}
\begin{tabular*}{0.3\textwidth}{@{\extracolsep{\fill}} c r }
\hline\hline
Particle         & Mass    \\ \hline
$h$              & $126$   \\
$A$              & $3168$  \\
$\chi^0_1$       & $405$   \\
$\chi^+_1$       & $626$   \\
$\tilde\nu_\tau$ & $1667$  \\
$\tilde\tau_1$   & $1666$  \\
$\tilde t_1$     & $4142$   \\
$\tilde b_1$     & $4583$   \\
\hline\hline
\end{tabular*}
\caption{Part of the supersymmetric spectrum (in GeV).
\label{tab2}}
\end{center}
\end{table}
This scenario was generated using the code SUSPECT \cite{Djouadi:2002ze} for the RpC part. In
particular, the Higgs boson mass is 126 GeV, as measured by experiments \cite{Aad:2012tfa}. In addition, SUSPECT 
allows the calculation for: (i) the deviation from unity of the $\rho$ parameter 
$\Delta\rho=7.7\times10^{-6}$ \cite{Beringer:1900zz,Ross:1975fq}, (ii) the anomalous magnetic moment 
of the muon $\Delta a_\mu=5.7\times10^{-11}$ \cite{Beringer:1900zz,Bennett:2006fi}, and (iii) the branching 
ratio for the radiative decay of a bottom quark $B(b\rightarrow s\gamma)=3.3\times10^{-4}$
\cite{Asner:2010qj}.

The BRpV part is handled by our own code. Since BRpV parameters are much smaller than the supersymmetric scale represented 
by the Higgsino mass parameter $\mu$, the extra contributions to the above loop quantities from BRpV are negligible.
\begin{table}
\begin{center}
\begin{tabular*}{0.4\textwidth}{@{\extracolsep{\fill}} c r c }
\hline\hline
Parameter    & Value     & Units        \\ \hline
$\epsilon_1$ & $0.162$   & GeV       \\
$\epsilon_2$ & $-0.043$   & GeV       \\
$\epsilon_3$ & $0.192$   & GeV       \\
$\Lambda_1$  & $0.153$  & GeV${}^2$ \\
$\Lambda_2$  & $0.178$  & GeV${}^2$ \\
$\Lambda_3$  & $0.067$  & GeV${}^2$ \\
\hline\hline
\end{tabular*}
\caption{BRpV parameters.
\label{tab3}}
\end{center}
\end{table}
The selected BRpV parameters are given in Table \ref{tab3}. Note that the values for $\epsilon_i$ are large enough 
to make the radiative corrections to neutrino masses very important. 
\begin{table}
\begin{center}
\begin{tabular*}{0.6\textwidth}{@{\extracolsep{\fill}} l l l c}
\hline\hline
Observable           & Central Value        & $3\sigma$ exp. value       & Units     \\ \hline
$\Delta m^2_{atm}$   & $2.56\times10^{-3}$  & $2.31-2.74 \times10 ^{-3}$ & eV${}^2$  \\
$\Delta m^2_{sol}$   & $7.62\times10^{-5}$  & $7.12-8.20 \times10 ^{-5}$ & eV${}^2$  \\
$\sin^2\theta_{atm}$ & $0.639$              & $0.36-0.68$                & -         \\
$\sin^2\theta_{sol}$ & $0.305$              & $0.27-0.37$                & -         \\
$\sin^2\theta_{rea}$ & $0.024$              & $0.017-0.033$              & -         \\
\hline\hline
\end{tabular*}
\caption{Experimental neutrino observables.
\label{tab4}}
\end{center}
\end{table}
The experimental values for the neutrino parameters are given in Table \ref{tab4}.

First of all, a study on how important are the different loops in the determination of the
neutrino parameters has been performed.
\begin{figure}[!ht*]
\begin{center}
\includegraphics[width=0.5\textwidth,angle=270]{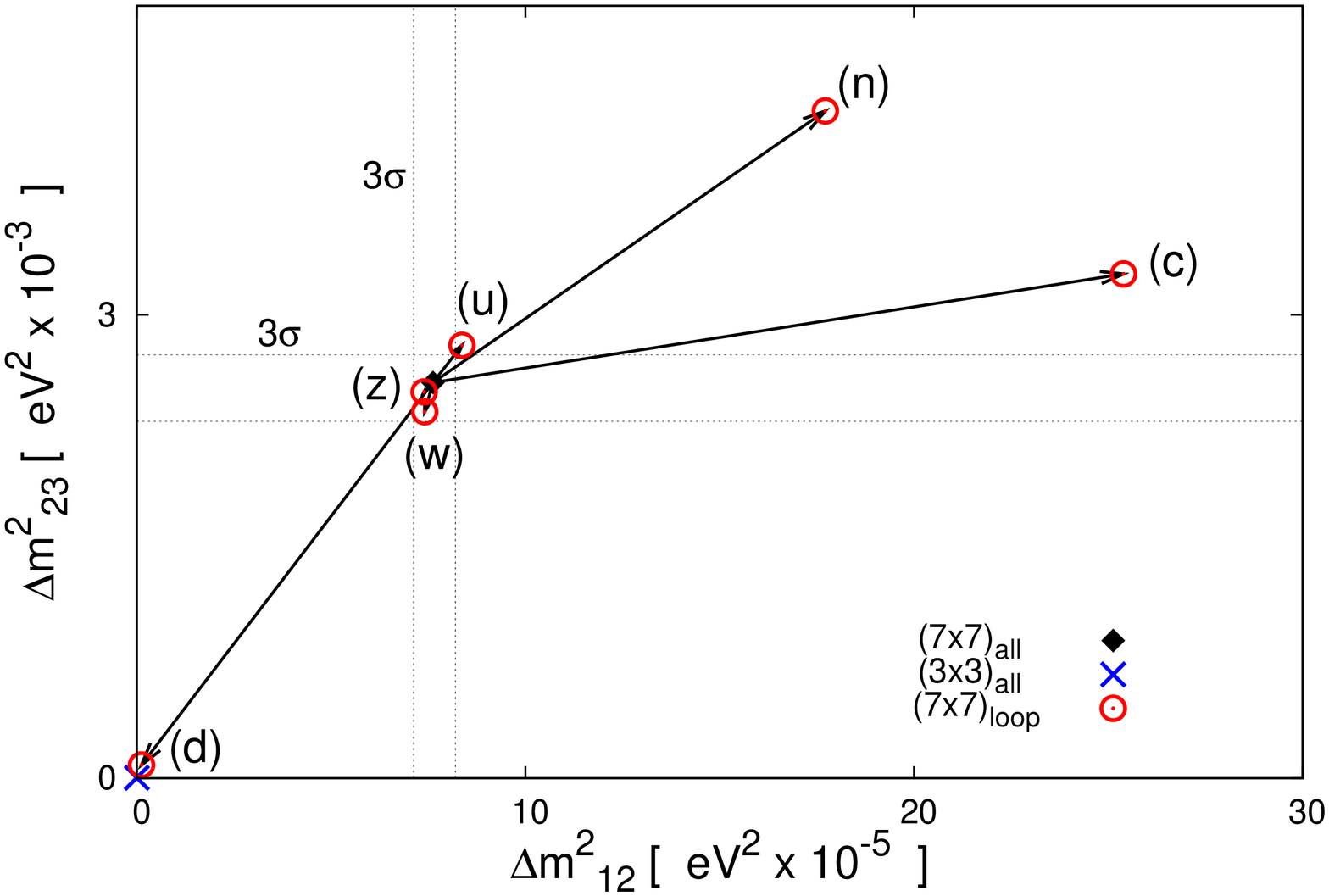}
\\
\includegraphics[width=0.5\textwidth,angle=270]{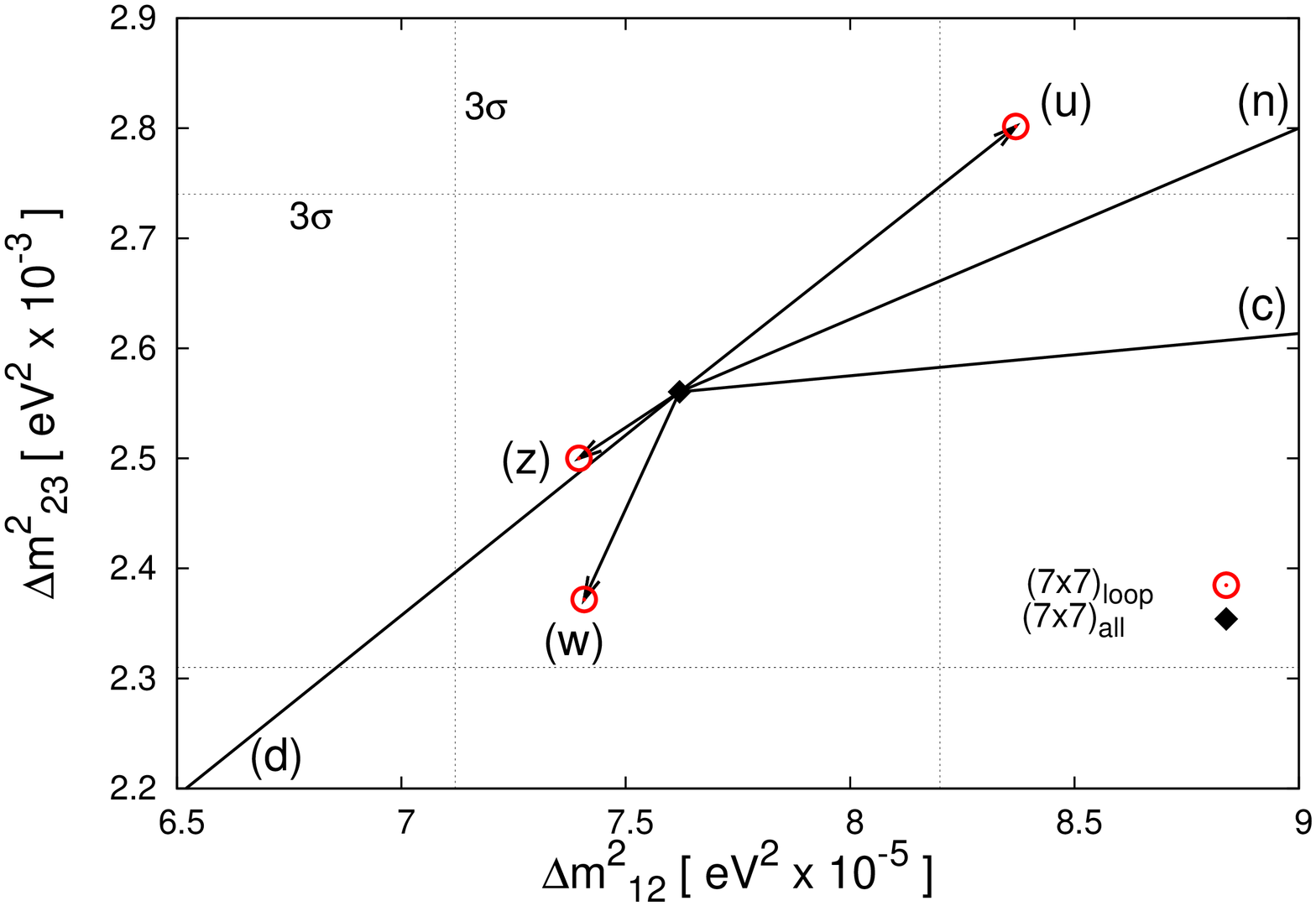}
\\
\caption{Influence of loop corrections on $\Delta m^2_{atm}$ and $\Delta m^2_{sol}$ in the 
whole $7\times 7$ mass matrix. The lower figure is a zoom-in of the top one.}
\label{p1loop_7x7_zout}
\end{center}
\end{figure}
In Fig.~\ref{p1loop_7x7_zout} one works in the plane formed by the atmospheric $\Delta m^2_{23}$ and the 
solar $\Delta m^2_{12}$ neutrino mass parameters. In vertical and horizontal dashed lines the $3\sigma$ 
experimental limits for these parameters are shown. At approximately the center of this allowed region one 
has the predictions from our scenario using the full $7\times7$ mass matrix, represented by a dark (black) 
diamond. Flowing from this point one has several arrows ending in circles (red), 
one for each loop. What it is done here is to omit in every entry of the $7\times7$ mass matrix the 
contribution from the corresponding loop, and show the prediction for the mass differences in these 
conditions.

The contributions from the bottom-sbottom, neutralino-neutral scalar, and chargino-charged scalar loops
are large as expected (Fig.~\ref{p1loop_7x7_zout}-top). The not-so-known
effect is the importance of the top-stop loops, which are large enough to move the prediction outside the
3-$\sigma$ region when it is not included (Fig.~\ref{p1loop_7x7_zout}-bottom). The reason for the
unexpectedness of this result is that these loops do not contribute to the neutrino masses in the $3\times3$
approach, which is very popular. The contribution by these loops appears through the last term in
eq.~(\ref{OneLoopCorr}), which is of third order. As explain in the next section, this contribution
is proportional to the top quark Yukawa coupling and needs the presence of the bottom-sbottom loops as well.
One may also see that in this particular scenario, the $3\times3$ approximation does not work since it gives a 
prediction for the solar and atmospheric mass squared parameters which are off by several orders of magnitude,
represented by a cross (blue).

\begin{figure}[!ht*]
\begin{center}
\includegraphics[width=0.5\textwidth,angle=270]{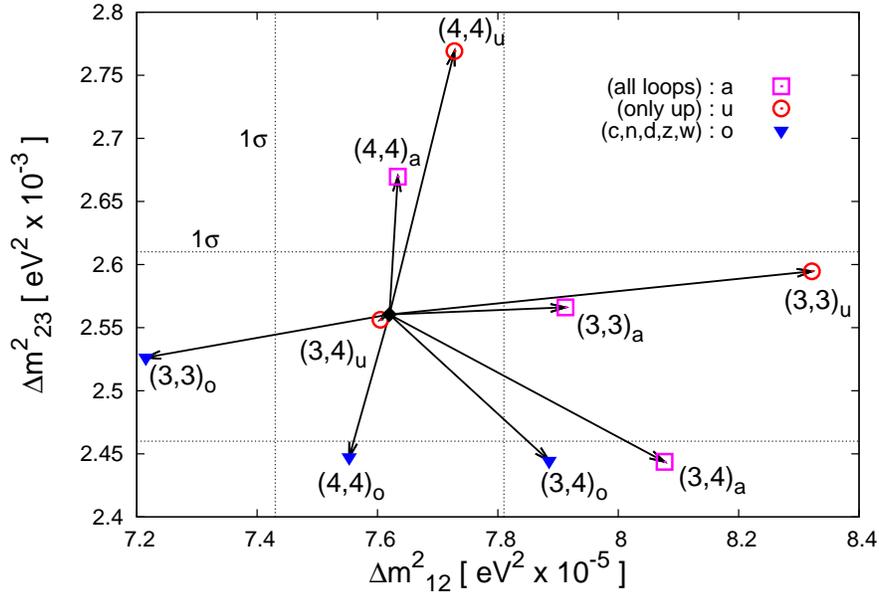}
\\
\caption{Influence of loop corrections on $\Delta m^2_{atm}$ and $\Delta m^2_{sol}$ in a
given matrix element of the $7\times 7$ mass matrix.}
\label{p1loop_neutralino}
\end{center}
\end{figure}
Second of all, in Fig.~\ref{p1loop_neutralino} a similar process is performed. 
This time a specific loop in a given entry in the $7\times7$ mass matrix is omitted. 
For the arrows ending in a square (magenta), one is omitting all the loops
at each $(3,3)$, $(3,4)$ and $(4,4)$ matrix elements. For the arrows ending in a 
circle (red), one is omitting the up-sup loops for the same matrix elements. 
Finally, for the arrows ending in a triangle (blue), one is omitting all the loops
except up-sup, also for the same matrix elements. The lesson draw from the figure is that 
the importance of the top-stop loops lies in the higgsino section of the mass matrix. This is clear 
since the corrections in that sector are proportional to the top quark Yukawa coupling.

When it is convenient to work with a $3\times3$ neutrino mass matrix,
the second and third order terms in eq.~(\ref{OneLoopCorr}) should be included, 
because they are numerically important.
Once that is done, the precision obtained with the $7\times7$ approach is recovered. 
\begin{table}
\begin{center}
\begin{tabular*}{0.6\textwidth}{@{\extracolsep{\fill}} l l l l}
\hline\hline
Observable           & $7\times7$          & $3\times3$          & ${3\times3}_{full}$ \\ \hline
$\Delta m^2_{atm}$   & $2.56\times10^{-3}$ & $2.02\times10^{-6}$ & $2.56\times10^{-3}$ \\
$\Delta m^2_{sol}$   & $7.62\times10^{-5}$ & $1.53\times10^{-8}$ & $7.57\times10^{-5}$ \\
$\sin^2\theta_{atm}$ & $0.639$             & $0.839$             & $0.640$             \\
$\sin^2\theta_{sol}$ & $0.305$             & $0.442$             & $0.303$             \\
$\sin^2\theta_{rea}$ & $0.024$             & $0.407$             & $0.024$             \\
\hline\hline
\end{tabular*}
\caption{Neutrino observables calculated in the different approaches.
\label{tab5}}
\end{center}
\end{table}
In Table \ref{tab5} the prediction for the neutrino observables in the same 
scenario introduced before is shown. In the second and third column the usual $7\times7$ and 
$3\times3$ approaches are shown. In the last column the extra terms in
eq.~(\ref{OneLoopCorr}), calling the approach as $3\times3_{full}$, is included. It is clear the 
recovery in precision.

The second order is given by the third term in eq.~(\ref{OneLoopCorr}). In the
chosen scenario, this term is also very important. That can be understood from 
Fig.~\ref{p1loop_7x7_zout}-top and Fig.~\ref{p1loop_neutralino}. In Fig.~\ref{p1loop_7x7_zout}-top
the effect of the first order is seen  by the cross (blue). The fact that this prediction
is so small is an indication that this first order effect is also small.
On the other hand, the effect of the third order seen in Fig.~\ref{p1loop_neutralino},
although large when compared to experimental errors, is small compared
to full expansion (first plus second plus third order), therefore, the
second order is very important.

%
\section{Algebraic Approximations} \label{sec:algebra}
%
Here, approximated algebraic expressions for second and third order terms
from the top-stop contribution to the solar mass are found, in order to 
better understand the numeric results shown in the previous section. 
These numerical calculations show that top-stop loops contribute importantly.

The contribution from top-stop loops to the second order term in 
eq.~(\ref{OneLoopCorr}) is studied. In the higgsino sector the relevant 
matrix elements of the inverse neutralino mass matrix, following the Appendix \ref{AppB} is,
\begin{equation}
({\mathrm{M}}_\chi^0)^{-1}_{34}  = ({\mathrm{M}}_\chi^0)^{-1}_{43} \approx -\frac{1}{\mu}.
\end{equation}
Therefore,
\begin{eqnarray}
- \left[ \delta m\,{\mathrm{M}}_{\chi^0}^{-1}\,\delta m^T \right]_{ij} &=&
\frac{1}{\mu} \Big[ \delta m_{i3} \delta m_{j4} + \delta m_{i4} \delta m_{j3} \Big]_{ij}
= \frac{1}{\mu} (\delta m_{3,\Lambda}^{t\tilde t}) (\delta m_{4,\Lambda}^{t\tilde t}) \, 
\Lambda_i\Lambda_j\,,
\end{eqnarray}
and it does not contribute to the solar mass, since it is proportional to 
$\Lambda_i \Lambda_j$. In fact, since the top-stop coupling to neutrinos does 
not include $\epsilon$ terms, none of the quantities $\delta m_{ij}^{t\tilde t}$ 
will produce a contribution to the solar mass. Thus, third order term is
studied next.

The third order term in eq.~(\ref{OneLoopCorr}), given by
\begin{equation}
\delta m \,({\mathrm M}^{d,0}_\chi)^{-1}\, \delta M_{\chi}\,
({\mathrm M}^{d,0}_\chi)^{-1} \delta m^T\,,
\label{Term1}
\end{equation}
is written in the basis where the tree-level neutralino mass matrix has already been diagonalized.
If work is to be done in the original basis instead, the term to analyze is,
\begin{equation}
\delta m \,({\mathrm M}^{0}_\chi)^{-1}\, \delta M_{\chi}\,
({\mathrm M}^{0}_\chi)^{-1} \delta m^T\,,
\label{Term2}
\end{equation}
where $\delta m$ (and $\delta M_{\tilde{\chi}}$) in eq.~(\ref{Term1}) is written in the diagonal 
basis, while $\delta m$ (and $\delta M_{\tilde{\chi}}$) in eq.~(\ref{Term2}) is written in the 
original basis. The same notation is used for both out of simplicity.

In order to algebraically understand the issues mentioned in the previous section a few approximations 
are performed. First, notice that down-type quarks contribute to $\delta m$ with a term proportional to 
$\epsilon_i$, while up-type quarks do not, as can be seen from the Appendix \ref{AppA}. Thus, in this approximation,
\begin{equation}
(\delta m)_{ij} = \delta m_{i3} \, \delta_{j3}.
\end{equation}
Second, notice that the $(4,4)$ matrix element in the neutralino sector makes a strong numerical 
effect on the neutrino parameters, and up-type quarks contribute to it. To isolate this effect 
it is assumed,
\begin{equation}
(\delta M_{\chi})_{ij} = \delta M_{\chi,44} \, \delta_{i4} \, \delta_{j4}.
\end{equation}
With this, the contribution from top-stop loops to the third order term 
in eq.~(\ref{OneLoopCorr}) is,
\begin{eqnarray}
\left[ \delta m\, ({\mathrm M}^{0}_\chi)^{-1} \, \delta M_{\chi} \, ({\mathrm M}^{0}_\chi)^{-1}
\,\delta m^T \right]_{ij}
&\approx&
\delta M_{\chi,44}^{t\tilde t} \, ({\mathrm M}^{0}_{\chi})^{-2}_{34}
(\delta m_{i3}^{b\tilde b}) \, (\delta m_{j3}^{b\tilde b})
\\ &\approx&
\delta M_{\chi,44}^{t\tilde t} \, ({\mathrm M}^{0}_{\chi})^{-2}_{34}
\Big[ \delta m_{3,\Lambda}^{b\tilde b} \Lambda_i + \delta m_{3,\epsilon}^{b\tilde b} \epsilon_i \Big]
\Big[ \delta m_{3,\Lambda}^{b\tilde b} \Lambda_j + \delta m_{3,\epsilon}^{b\tilde b} \epsilon_j \Big].
\nonumber
\end{eqnarray}
Approximating further the $\epsilon\epsilon$ term is,
\begin{eqnarray}
\left[ \delta m\, ({\mathrm M}^{0}_\chi)^{-1} \, \delta M_{\chi} \, ({\mathrm M}^{0}_\chi)^{-1}
\,\delta m^T \right]_{ij}^{\epsilon\epsilon}
&\approx& 
\left[ \frac{n_ch_t^2}{32\pi^2}\times 2 \times 2\mu \right] \, \left[ -\frac{1}{\mu} \right]^2 \, \left[
\frac{n_ch_b^2}{64\pi^2\mu}\times 2 \times 2\mu \right]^2 \epsilon_i\epsilon_j 
\nonumber\\
&=& 
\frac{2n_c^3h_t^2 h_b^4}{(16\pi^2)^3\mu}\epsilon_i\epsilon_j
\approx
\frac{n_c^3 g^6m_t^2m_b^4}{4(16\pi^2)^3s_\beta^2c_\beta^4m_W^6\mu}\epsilon_i\epsilon_j
\nonumber\\
&\approx& 
10^{-2} \frac{t_\beta^4\epsilon_i\epsilon_j}{\mu} \,\,{\text{eV}}\,,
\end{eqnarray}
which gives the same order of magnitude of the solar mass squared difference, thus it should not
be neglected.

%
\section{General Scan over Parameter Space}
%

In order to see the 0 of the different approximations a general scan over the parameter space was made.
As it was explained in section \ref{sec:HOE} the code SUSPECT \cite{Djouadi:2002ze} was used for the running of the 
RpC supersymmetric parameters, and our own code for the neutrino observables from the R-Parity 
violating parameters (since R-Parity violation is small, the use of MSSM RGEs for the RpC parameters is a good 
approximation). Randomly selected values for the RpC parameters at the GUT scale are generated and SUSPECT is used 
to find their counterpart at the weak scale, including a correct electroweak symmetry breaking. At this point 
the following cuts are added: the Higgs mass, $124<m_h<126$ GeV, $\Delta\rho$, $\Delta a_\mu$, 
$B(b\rightarrow s\gamma)$ (see first paragraph in section 3). Then, randomly generated values for the RpV parameters
are added to the 0 parameters, and with all of them a seed point in parameter space at the weak scale is
defined. Using an implementation of the Markov chain \cite{Numerics} and starting from the seed point just described,
a movement in steps is implemented, minimizing a $\chi^2$ function that includes neutrino experimental parameters only 
(mass squared differences and mixing angles) towards a final point that satisfy neutrino physics. Finally, cuts on 
the masses of the following supersymmetric particles are implemented $m_{\chi^+_1}>600$ GeV, $m_{\chi^0_1}>300$ GeV, 
$m_{\tilde \ell}>1$ TeV, $m_{\tilde q}>2$ TeV, $m_{\tilde g}>2$ TeV \cite{Aad:2014pda,CMS:yut}

Following section \ref{sec:HOE}, some of the parameters at the weak scale that define the points that satisfy
all cuts lie in intervals described in Table \ref{tab6}.
\begin{table}
\begin{center}
\begin{tabular*}{0.5\textwidth}{@{\extracolsep{\fill}} c c c c }
\hline\hline
Parameter    & Minimum  & Maximum  & Units     \\ \hline
$\tan\beta$  & $9.11$   & $48.8$   &  -        \\
$\mu$        & $655$    & $4495$   & GeV       \\
$M_1$        & $313$    & $897$    & GeV       \\
$M_2$        & $567$    & $1597$   & GeV       \\
$M_3$        & $3296$   & $5952$   & GeV       \\
$M_Q$        & $2862$   & $6093$   & GeV       \\
$M_U$        & $1616$   & $5834$   & GeV       \\
$M_D$        & $2427$   & $6458$   & GeV       \\
$M_L$        & $1007$   & $5176$   & GeV       \\
$M_R$        & $1024$   & $4899$   & GeV       \\
$\epsilon_1$ & $-0.117$ & $0.158$  & GeV       \\
$\epsilon_2$ & $-0.235$ & $0.303$  & GeV       \\
$\epsilon_3$ & $-0.156$ & $0.277$  & GeV       \\
$\Lambda_1$  & $-0.102$ & $0.112$  & GeV${}^2$ \\
$\Lambda_2$  & $-0.118$ & $0.124$  & GeV${}^2$ \\
$\Lambda_3$  & $-0.130$ & $0.116$  & GeV${}^2$ \\
\hline\hline
\end{tabular*}
\caption{Intervals for each parameters at the low scale.
\label{tab6}}
\end{center}
\end{table}
Similarly, the interval for some of the physical masses are indicated in Table \ref{tab7}.
\begin{table}
\begin{center}
\begin{tabular*}{0.4\textwidth}{@{\extracolsep{\fill}} c c c }
\hline\hline
Particle         & Minimum & Maximum \\ \hline
$h$              & $124$   & $126$   \\
$A$              & $1097$  & $4128$  \\
$\chi^0_1$       & $310$   & $897$   \\
$\chi^+_1$       & $601$   & $1651$  \\
$\tilde\nu_\tau$ & $1005$  & $5176$  \\
$\tilde\tau_1$   & $1001$  & $4800$  \\
$\tilde t_1$     & $2973$  & $6234$  \\
$\tilde b_1$     & $3216$  & $6559$  \\
\hline\hline
\end{tabular*}
\caption{Part of the supersymmetric spectrum (in GeV).
\label{tab7}}
\end{center}
\end{table}
The high value for the squark masses (and soft mass parameters as well) is due to the Higgs mass, which needs 
high squarks masses in order to reach the value $m_h\sim 125$ GeV. For the same reason (although contributing 
at two loops), the gluino mass is also high: $m_{\tilde g}>3500$ GeV including radiative corrections.

\begin{figure}[!ht*]
\begin{center}
\includegraphics[width=0.4\textwidth,angle=270]{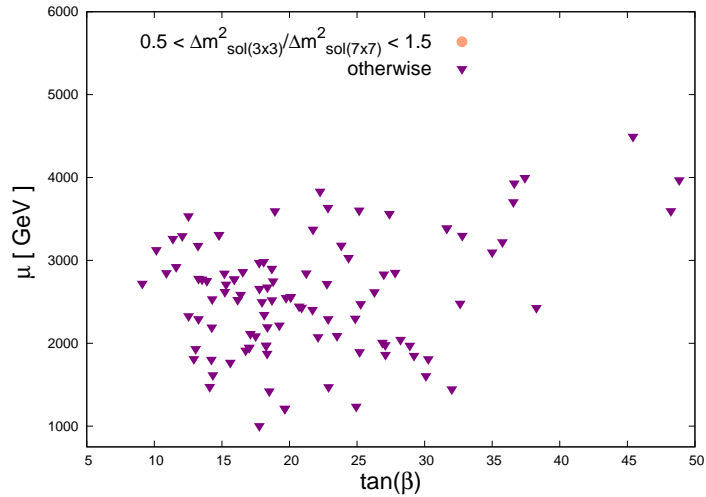}
\\
\caption{Solar mass squared difference calculated with the $3\times3$ approximation, in comparison with the one
calculated with the $7\times7$ matrix.}
\label{scan3x3}
\end{center}
\end{figure}
In Fig.~\ref{scan3x3} the 0 of the $3\times3$ approximation in the $\mu$-$\tan\beta$ plane is shown. 
Different colors according to the ratio $\Delta m^2_{sol (3\times3)}/\Delta m^2_{sol (7\times7)}$ are displayed
(in principle), {\sl i.e.}, the solar mass squared difference calculated with the $3\times3$ approximation in 
comparison with the same neutrino observable calculated with the full $7\times7$ matrix. It is seen that the solar 
mass calculated with the $3\times3$ approximation is always off by more than $50\%$. In fact, it is observed in 
the scan that it is always smaller, and very often the error is much larger than $50\%$. Considering the experimental 
errors in the measurement of the solar mass, the $3\times3$ approximation is not reliable anymore.

\begin{figure}[!ht*]
\begin{center}
\includegraphics[width=0.5\textwidth,angle=270]{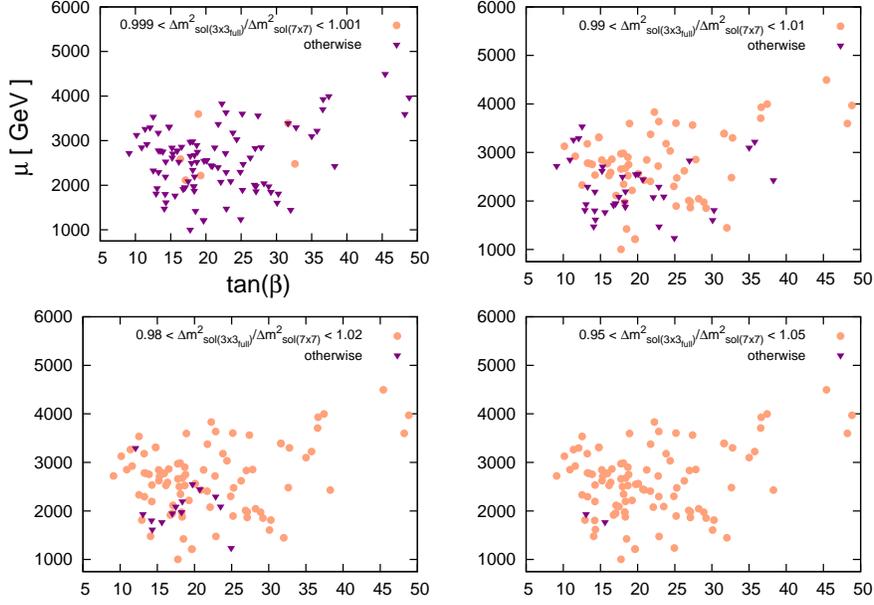}
\\
\caption{Solar mass squared difference calculated with the $3\times3_{full}$ approximation, in comparison with the one
calculated with the $7\times7$ matrix.}
\label{composite2}
\end{center}
\end{figure}
In Fig.~\ref{composite2} a similar comparison is made, but this time for the solar mass calculated with the 
$3\times3_{full}$ approximation. Furthermore, displayed are four quadrants that refer to four different values for 
the error: $0.1\%$, $1\%$, $2\%$, and $5\%$. In the lower-right frame ($5\%$) it is seen that the $3\times3_{full}$
approximation is much better than the usual $3\times3$: almost all the time the solar mass lies within $5\%$ with 
respect to the calculated with the $7\times7$ matrix. At the level of $0.1\%$ (upper-left), even the $3\times3_{full}$ 
approximation fails in comparison with $7\times7$. In addition, from the sequence of frames in Fig.~\ref{composite2}
it can be concluded that the $3\times3_{full}$ is more reliable at high values of $\tan\beta$. This can be understood 
from the fact that at high values of $\tan\beta$ the bottom quark Yukawa coupling is larger and, therefore, bottom quark 
effects are more 0. This makes the $3\times3_{full}$ approximation more reliable, and simultaneously the 
$3\times3$ approximation less reliable, at high values of $\tan\beta$.

\begin{figure}[!ht*]
\begin{center}
\includegraphics[width=0.5\textwidth,angle=270]{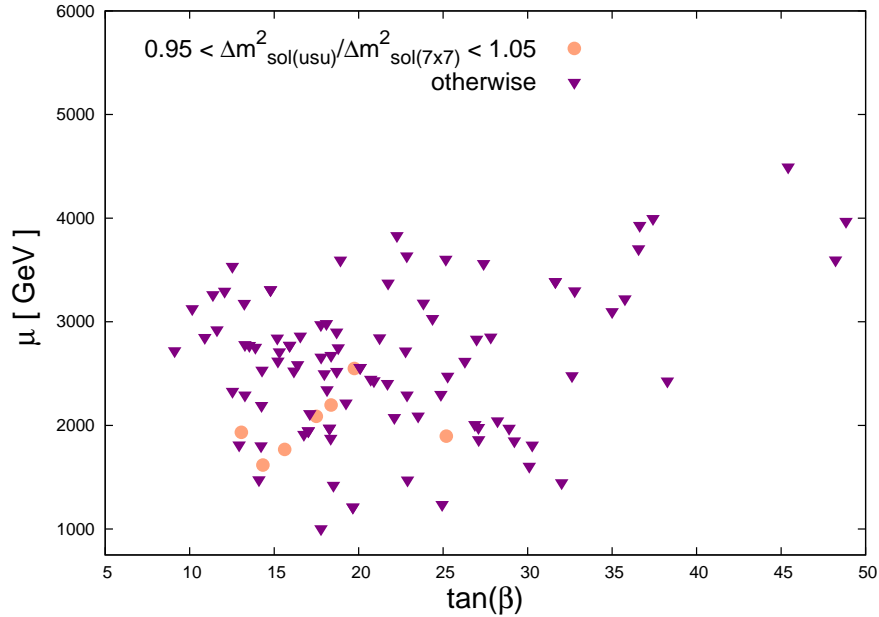}
\\
\caption{Effect of the removal of the up-type quark and squark loops.}
\label{scanusu}
\end{center}
\end{figure}
Finally in Fig.~\ref{scanusu} it is seen the effect of the up type quarks and squarks on the solar mass, displayed in 
the same $\mu-\tan\beta$ plane. Notice that these loops contribute to the solar mass only via the third order term. 
In the scan the effect of removing all together the up-sup loops from the $7\times7$ matrix is shown, and a
0 with the solar mass calculated with the full $7\times7$ matrix is done. In most of the points the effect 
of the up-sup loops is large (larger than $5\%$ in the figure).

\section{Conclusions}

It was shown that the $3\times3$ approach in the calculation of neutrino masses in the MSSM with BRpV, 
in the light of the present accuracy of the experimental results, sometimes does not give an acceptable
answer. This was understood by studying the $3\times3$ second and third order terms in the block 
diagonalization of the $7\times7$ mass matrix. When it is convenient to work with $3\times3$ matrices, 
it was shown also that keeping these terms gives a very similar result 
compared to the ones extracted from the $7\times7$ neutrino mass matrix. In addition, in the 
$3\times3$ approach, the top-stop loops do not contribute, nevertheless, they can be numerically 
important. These loops contribute through the already mentioned third order term, and it was shown that 
the contribution is dependent on the bottom as well as the top quark Yukawa couplings.
The second order term in eq.~(\ref{OneLoopCorr}) can also be very important. In fact,
a scenario was chosen where it is crucial. All these issues motivate a two-loop calculation
of neutrino masses in this model. A scan over parameter space is made to show that the conclusions are general,
and not driven by the chosen point shown in section 3. Most of these numerical effect come from the high value 
of the Higgs boson mass.

\begin{center}
\textbf{Acknowledgments} 
\end{center}

This work was supported by Fondecyt grants No. 11110472, 1100837 and 1141190, Anillo ``AtlasAndino'' 
ACT1102, UTFSM-DGIP grant No. 11.12.39, and Conicyt Doctorate Grant.


\appendix
\section{Squark Loop Contributions}\label{AppA}

\subsection{Top-stop loops in $\delta M_\chi$}

It is numerically observed that among the 16 matrix elements of $\delta M_\chi$, the $(4,4)$ 
is the one that gives the largest contribution. In addition, the top-stop loops have an 
important effect on this matrix element. In order to algebraically understand the phenomena, 
this contribution is calculated. The coupling between neutral fermions and top-stop quarks is,
\begin{center}
\vspace{-50pt} \hfill \\
\begin{picture}(110,90)(0,23) 
\DashLine(10,25)(50,25){4}
\ArrowLine(50,25)(78,53)
\ArrowLine(78,-3)(50,25)
\Text(10,35)[]{$\tilde t_k$}
\Text(90,55)[]{$t$}
\Text(90,-5)[]{$F^0_j$}
\end{picture}
$
=\,i\,\Big[O^{tn\tilde t}_{Ljk}\frac{(1-\gamma_5)}{2}+
O^{tn\tilde t}_{Rjk}\frac{(1+\gamma_5)}{2}\Big],
$
\vspace{30pt} \hfill \\
\end{center}
\vspace{10pt}
with
\begin{eqnarray}
O^{tn\tilde t}_{Ljk} &=& \eta_j \frac{4gt_W}{3\sqrt{2}} {\cal N}^*_{j1} R^{\tilde t}_{k2} - 
\eta_j h_t {\cal N}^*_{j4} R^{\tilde t}_{k1}\,,
\nonumber\\
O^{tn\tilde t}_{Rjk} &=& -\frac{g}{\sqrt{2}} \left({\cal N}_{j2}+\frac{1}{3}t_W { \cal N}_{j1} \right) 
R^{\tilde t}_{k1} - h_t {\cal N}_{j4} R^{\tilde t}_{k2}\,,
\label{F0topstopCoup}
\end{eqnarray}
and where $h_t$ is the top quark Yukawa coupling, $R^{\tilde t}$ is the (assumed real) $2\times2$ 
rotation matrix that diagonalizes the stop quark mass matrix, ${\cal N}$ is the (assumed real) $7\times7$ 
rotation matrix that diagonalizes the neutralino sector, and $\eta_j$ is the sign of the 
corresponding fermion $j$. Notice that the complex conjugated ${\cal N}^*$ is kept only for 
reference, since one assumes it is real. If this coupling is specialized to the case when the neutral 
fermion is a neutralino one finds,
\begin{center}
\vspace{-50pt} \hfill \\
\begin{picture}(110,90)(0,23) 
\DashLine(10,25)(50,25){4}
\ArrowLine(50,25)(78,53)
\ArrowLine(78,-3)(50,25)
\Text(10,35)[]{$\tilde t_k$}
\Text(90,55)[]{$t$}
\Text(90,-5)[]{$\chi^0_j$}
\end{picture}
$
=\,i\,\Big[O^{t\chi\tilde t}_{Ljk}\frac{(1-\gamma_5)}{2}+
O^{t\chi\tilde t}_{Rjk}\frac{(1+\gamma_5)}{2}\Big],
$
\vspace{30pt} \hfill \\
\end{center}
\vspace{10pt}
with
\begin{eqnarray}
O^{t\chi\tilde t}_{Ljk} &=& \eta_j \frac{4gt_W}{3\sqrt{2}} N^*_{j1} R^{\tilde t}_{k2} - 
\eta_j h_t N^*_{j4} R^{\tilde t}_{k1}\,,
\nonumber\\
O^{t\chi\tilde t}_{Rjk} &=& -\frac{g}{\sqrt{2}} \left(N_{j2}+\frac{1}{3}t_W N_{j1} \right) R^{\tilde t}_{k1} 
- h_t N_{j4} R^{\tilde t}_{k2}\,.
\label{X0topstopCoup}
\end{eqnarray}
In this case, $N$ is the (real) $4\times4$ rotation matrix that diagonalizes the neutralino mass 
sub-matrix, and $\eta_j$ is the sign of the $j$-th neutralino mass.
The relevant loop is formed with those couplings,
\begin{center}
\vspace{-50pt} \hfill \\
\begin{picture}(110,90)(0,23) 
\ArrowLine(0,25)(30,25)
\ArrowArc(50,25)(20,180,0)
\ArrowLine(70,25)(100,25)
\DashCArc(50,25)(20,0,180){4}
\Text(51,53)[]{$\tilde t_k$}
\Text(50,-4)[]{$t$}
\Text(10,35)[]{$\chi^0_i$}
\Text(90,35)[]{$\chi^0_j$}
\end{picture}
$
=\,i\,\Sigma_{ij}^{t\tilde t}(p^2),
$
\vspace{30pt} \hfill \\
\end{center}
\vspace{10pt}
with,
\begin{equation}
\Sigma_{ij}^{t\tilde t}(p^2) = \frac{n_ch_t^2 N_{i4} N_{j4}}{16\pi^2} \sum_{k=1}^2 \left[
m_t R^{\tilde t}_{k1} R^{\tilde t}_{k2} (\eta_iP_L+\eta_jP_R) B_0^{pt\tilde t} -
\slashed{p} (R^{\tilde t 2}_{k1} \eta_i\eta_j P_L + R^{\tilde t 2}_{k2} P_R) B_1^{pt\tilde t}
\right]+...
\end{equation}
Here the three dots mean that only the terms proportional to $h_t^2$ are shown. Also, the fact
that the matrix $N$ is real was already used.

When evaluating $\delta M_\chi^{ij}$ it should be understood that in the basis where the neutralinos 
are diagonal, one wants to evaluate the neutralino mass at $p^2$, and symmetrize 
over $i$ and $j$. Thus,
\begin{eqnarray}
\delta M_\chi^{ij} &=& \frac{n_ch_t^2 N_{i4} N_{j4}}{32\pi^2} \sum_{k=1}^2 \bigg\{ - \frac{1}{2} m_t
R^{\tilde t}_{k1} R^{\tilde t}_{k2} (\eta_i+\eta_j) 
\left( B_0^{\chi_i t\tilde t}+B_0^{\chi_j t\tilde t} \right)
\nonumber\\
&& \qquad\qquad\qquad\qquad
+ \frac{1}{2} (R^{\tilde t 2}_{k1}\eta_i\eta_j+R^{\tilde t 2}_{k2}) \left( m_{\chi^0_i} 
B_1^{\chi_i t \tilde t} + m_{\chi^0_j} B_1^{\chi_j t \tilde t} \right) \bigg\}\,,
\nonumber\\
&=&
\frac{n_ch_t^2 N_{i4} N_{j4}}{32\pi^2} \sum_{k=1}^2 \bigg\{ \frac{1}{2} m_t
s_{\tilde t} c_{\tilde t} (-1)^k (\eta_i+\eta_j) 
\left( B_0^{\chi_i t\tilde t}+B_0^{\chi_j t\tilde t} \right)
\nonumber\\
&& \qquad\qquad\qquad\qquad
+ \frac{1}{2} (c_{\tilde t}^2\eta_i\eta_j+s_{\tilde t}^2) \left( m_{\chi^0_i} 
B_1^{\chi_i t \tilde t} + m_{\chi^0_j} B_1^{\chi_j t \tilde t} \right) \bigg\}\,.
\end{eqnarray}
The contribution to the $(4,4)$ neutrino/neutralino mass matrix element is therefore,
\begin{equation}
\delta M_\chi^{44} =
\frac{n_ch_t^2 N_{44}^2}{32\pi^2} \sum_{k=1}^2 \bigg\{ 2m_t s_{\tilde t} c_{\tilde t} (-1)^k
\eta_4 B_0(m_{\chi^0_4}^2; m_t^2, m_{\tilde t_k}^2)
+ m_{\chi^0_4} B_1(m_{\chi^0_4}^2; m_t^2, m_{\tilde t_k}^2) \bigg\},
\end{equation}
which is an approximation for the top-stop loop contribution to $\delta M_\chi^{44}$.

\subsection{Bottom-sbottom loops in $\delta M_\chi$}

Bottom-sbottom loops contribute importantly to $\delta M_\chi$, and through it, also contribute 
importantly to the third term in eq.~(\ref{OneLoopCorr}). Bottom-sbottom 
loops contribute importantly to $\delta M_{\nu}$ too, but they are not the focus of 
this study. The neutral fermion coupling to bottom-sbottom quarks is,
\begin{center}
\vspace{-50pt} \hfill \\
\begin{picture}(110,90)(0,23) 
\DashLine(10,25)(50,25){4}
\ArrowLine(50,25)(78,53)
\ArrowLine(78,-3)(50,25)
\Text(10,35)[]{$\tilde b_k$}
\Text(90,55)[]{$b$}
\Text(90,-5)[]{$F^0_j$}
\end{picture}
$
=\,i\,\Big[O^{bn\tilde b}_{Ljk}\frac{(1-\gamma_5)}{2}+
O^{bn\tilde b}_{Rjk}\frac{(1+\gamma_5)}{2}\Big],
$
\vspace{30pt} \hfill \\
\end{center}
\vspace{10pt}
with
\begin{eqnarray}
O^{bn\tilde b}_{Ljk} &=&
- \eta_j \frac{2gt_W}{3\sqrt{2}} 
{\cal N}_{j1}^* R^{\tilde b}_{k2} - \eta_j h_b {\cal N}_{j3}^* R^{\tilde b}_{k1}\,,
\nonumber\\
O^{bn\tilde b}_{Rjk} &=& 
\frac{g}{\sqrt{2}} \left( {\cal N}_{j2} - \frac{1}{3} t_W {\cal N}_{j1} \right) R^{\tilde b}_{k1}
- h_b {\cal N}_{j3} R^{\tilde b}_{k2}\,,
\label{F0bottomsbottomCoup}
\end{eqnarray}
and where $h_b$ is the bottom quark Yukawa coupling, $R^{\tilde b}$ is the (assumed real) $2\times2$ 
rotation matrix that diagonalizes the sbottom quark mass matrix, ${\cal N}$ is the already defined 
(and real) $7\times7$ rotation matrix that diagonalizes the neutralino sector, and $\eta_j$ is 
the already defined sign of the corresponding fermion $j$. Specializing this coupling to the 
case when the neutral fermion is a neutralino, one finds,
\begin{center}
\vspace{-50pt} \hfill \\
\begin{picture}(110,90)(0,23) 
\DashLine(10,25)(50,25){4}
\ArrowLine(50,25)(78,53)
\ArrowLine(78,-3)(50,25)
\Text(10,35)[]{$\tilde b_k$}
\Text(90,55)[]{$b$}
\Text(90,-5)[]{$\chi^0_j$}
\end{picture}
$
=\,i\,\Big[O^{b\chi\tilde b}_{Ljk}\frac{(1-\gamma_5)}{2}+
O^{b\chi\tilde b}_{Rjk}\frac{(1+\gamma_5)}{2}\Big],
$
\vspace{30pt} \hfill \\
\end{center}
\vspace{10pt}
with
\begin{eqnarray}
O^{b\chi\tilde b}_{Ljk} &=&
- \eta_j \frac{2gt_W}{3\sqrt{2}} 
N_{j1}^* R^{\tilde b}_{k2} - \eta_j h_b N_{j3}^* R^{\tilde b}_{k1}\,,
\nonumber\\
O^{b\chi\tilde b}_{Rjk} &=& 
\frac{g}{\sqrt{2}} \left( N_{j2} - \frac{1}{3} t_W N_{j1} \right) R^{\tilde b}_{k1}
- h_b N_{j3} R^{\tilde b}_{k2}\,,
\label{X0bottomsbottomCoup}
\end{eqnarray}
and where $N$ is the already defined (real) $4\times4$ rotation matrix that diagonalizes the 
neutralino mass sub-matrix. The bottom-sbottom loops are,
\begin{center}
\vspace{-50pt} \hfill \\
\begin{picture}(110,90)(0,23) 
\ArrowLine(0,25)(30,25)
\ArrowArc(50,25)(20,180,0)
\ArrowLine(70,25)(100,25)
\DashCArc(50,25)(20,0,180){4}
\Text(51,53)[]{$\tilde b_k$}
\Text(50,-4)[]{$b$}
\Text(10,35)[]{$\chi^0_i$}
\Text(90,35)[]{$\chi^0_j$}
\end{picture}
$
=\,i\,\Sigma_{ij}^{b\tilde b}(p^2),
$
\vspace{30pt} \hfill \\
\end{center}
\vspace{10pt}
where
\begin{equation}
\Sigma_{ij}^{b\tilde b}(p^2) = \frac{n_ch_b^2N_{i3}N_{j3}}{16\pi^2} \sum_{k=1}^2 \left[
m_b R^{\tilde b}_{k1} R^{\tilde b}_{k2} (\eta_iP_L+\eta_jP_R) B_0^{pb\tilde b} -
\slashed{p} (R^{\tilde b 2}_{k1}\eta_i\eta_j P_L + R^{\tilde b 2}_{k2} P_R) B_1^{pb\tilde b}
\right]+...
\end{equation}
and again, only the terms proportional to $h_b^2$ are shown. The contribution to
$\delta M_\chi^{ij}$ is therefore,
\begin{eqnarray}
\delta M_\chi^{ij} &=& \frac{n_ch_b^2 N_{i3} N_{j3}}{32\pi^2} \sum_{k=1}^2 \bigg\{ 
- \frac{1}{2} m_b R^{\tilde b}_{k1} R^{\tilde b}_{k2} (\eta_i+\eta_j) 
\left( B_0^{\chi_i b\tilde b}+B_0^{\chi_j b\tilde b} \right)
\nonumber\\
&& \qquad\qquad\qquad\qquad
+ \frac{1}{2} (R^{\tilde b 2}_{k1}\eta_i\eta_j+R^{\tilde b 2}_{k2}) \left( m_{\chi^0_i} 
B_1^{\chi_i b \tilde b} + m_{\chi^0_j} B_1^{\chi_j b \tilde b} \right) \bigg\},
\nonumber\\
&=&
\frac{n_ch_b^2 N_{i3} N_{j3}}{32\pi^2} \sum_{k=1}^2 \bigg\{ \frac{1}{2} m_b
s_{\tilde b} c_{\tilde b} (-1)^k (\eta_i+\eta_j) 
\left( B_0^{\chi_i b\tilde b}+B_0^{\chi_j b\tilde b} \right)
\nonumber\\
&& \qquad\qquad\qquad\qquad
+ \frac{1}{2} (c_{\tilde b}^2\eta_i\eta_j+s_{\tilde b}^2) \left( m_{\chi^0_i} 
B_1^{\chi_i b \tilde b} + m_{\chi^0_j} B_1^{\chi_j b \tilde b} \right) \bigg\}.
\end{eqnarray}
The contribution to the $(4,4)$ neutrino/neutralino mass matrix element is therefore,
\begin{equation}
\delta M_\chi^{44} =
\frac{n_ch_b^2 N_{43}^2}{32\pi^2} \sum_{k=1}^2 \bigg\{ 2m_b s_{\tilde b} c_{\tilde b} (-1)^k
\eta_4 B_0(m_{\chi^0_4}^2; m_b^2, m_{\tilde b_k}^2)
+ m_{\chi^0_4} B_1(m_{\chi^0_4}^2; m_b^2, m_{\tilde b_k}^2) \bigg\},
\end{equation}
which is an approximation for the bottom-sbottom loop contribution to $\delta M_\chi^{44}$.

\subsection{Top-stop loops in $\delta m$}

In $\delta m$ one has mixing between neutralinos and neutrinos. Therefore, to find the quantum 
corrections in this region of the mass matrix the neutralino-top-stop coupling in 
eq.~(\ref{X0topstopCoup}) is needed. Also, to specialize the general coupling in 
eq.~(\ref{F0topstopCoup}) to the neutrino-top-stop coupling is needed. One finds,
\begin{center}
\vspace{-50pt} \hfill \\
\begin{picture}(110,90)(0,23) 
\DashLine(10,25)(50,25){4}
\ArrowLine(50,25)(78,53)
\ArrowLine(78,-3)(50,25)
\Text(10,35)[]{$\tilde t_k$}
\Text(90,55)[]{$t$}
\Text(90,-5)[]{$\nu_j$}
\end{picture}
$
=\,i\,\Big[O^{t\nu\tilde t}_{Ljk}\frac{(1-\gamma_5)}{2}+
O^{t\nu\tilde t}_{Rjk}\frac{(1+\gamma_5)}{2}\Big],
$
\vspace{30pt} \hfill \\
\end{center}
\vspace{10pt}
with
\begin{eqnarray}
O^{t\nu\tilde t}_{Ljk} &=& \eta_j R^{\tilde t}_{k1} h_t \xi_{j4}
- \eta_j R^{\tilde t}_{k2} \frac{4gt_W}{3\sqrt{2}} \xi_{j1}
\equiv \eta_j \widetilde O^{t\nu\tilde t}_{Lk} \, \Lambda_j\,,
\nonumber\\
O^{t\nu\tilde t}_{Rjk} &=& R^{\tilde t}_{k1} \frac{g}{\sqrt{2}}
\left( \xi_{j2} + \frac{1}{3} t_W \xi_{j1} \right)
+ R^{\tilde t}_{k2} h_t \xi_{j4}
\equiv \widetilde O^{t\nu\tilde t}_{Rk} \, \Lambda_j\,.
\end{eqnarray}
In the last equalities, the $\widetilde O$ couplings are defined as,
\begin{eqnarray}
\widetilde O^{t\nu\tilde t}_{Lk} &=& R^{\tilde t}_{k1} h_t \xi_4
- R^{\tilde t}_{k2} \frac{4gt_W}{3\sqrt{2}} \xi_1\,,
\nonumber\\
\widetilde O^{t\nu\tilde t}_{Rk} &=& R^{\tilde t}_{k1} \frac{g}{\sqrt{2}}
\left( \xi_2 + \frac{1}{3} t_W \xi_1 \right)
+ R^{\tilde t}_{k2} h_t \xi_4\,,
\end{eqnarray}
and the $\xi_{ij}$ and $\xi_i$ parameters are defined in Appendix \ref{AppC}.
The loops contributing to $\delta m$ are,
\begin{center}
\vspace{-50pt} \hfill \\
\begin{picture}(110,90)(0,23) 
\ArrowLine(0,25)(30,25)
\ArrowArc(50,25)(20,180,0)
\ArrowLine(70,25)(100,25)
\DashCArc(50,25)(20,0,180){4}
\Text(51,53)[]{$\tilde t_k$}
\Text(50,-4)[]{$t$}
\Text(10,35)[]{$\nu_i$}
\Text(90,35)[]{$\chi^0_j$}
\end{picture}
$
=\,i\,\Sigma_{i+4,j}^{t\tilde t}(p^2),
$
\vspace{30pt} \hfill \\
\end{center}
\vspace{10pt}
with
\begin{equation}
\Sigma_{i+4,j}^{t\tilde t}(p^2) = -\frac{n_ch_t^2\xi_{i4}N_{j4}}{16\pi^2} \sum_{k=1}^2 \left[
m_t R^{\tilde t}_{k1} R^{\tilde t}_{k2} (\eta_iP_L+\eta_jP_R) B_0^{pt\tilde t} -
\slashed{p} (R^{\tilde t 2}_{k1}\eta_i\eta_j P_L + R^{\tilde t 2}_{k2} P_R) B_1^{pt\tilde t}
\right]+...
\end{equation}
where only terms proportional to the Yukawa coupling squared are kept.
The above leads to the following contribution to $\delta m$,
\begin{eqnarray}
\delta m_{ij}^{t\tilde t} &=& \frac{n_ch_t^2\xi_4N_{j4}}{64\pi^2} \sum_{k=1}^2 \bigg\{
m_t s_{\tilde t} c_{\tilde t} (-1)^k \left( \eta_i+\eta_j \right) \Big[
B_0(m_{\chi_j^0}^2;m_t,m_{\tilde t_k}) + B_0(0;m_t,m_{\tilde t_k}) \Big]
\nonumber\\ && \qquad\qquad\quad
- m_{\chi_4^0} \left( \eta_i\eta_jc_{\tilde t}^2 + s_{\tilde t}^2 \right) 
B_1(m_{\chi_j^0}^2;m_t,m_{\tilde t_k})
\bigg\} \Lambda_i\,.
\label{dmTopStop}
\end{eqnarray}
From this result one learns that the second order term in eq.~(\ref{OneLoopCorr}) will never generate 
a solar neutrino mass from top-stop loops. This last conclusion arises because there is no term proportional to $\epsilon_i$ in eq.~(\ref{dmTopStop}).

\subsection{Bottom-sbottom loops in $\delta m$}

As it was mentioned before, in $\delta m$ one has mixing between neutralinos and neutrinos. 
The contribution from bottom-sbottom loops to this quantity starts with the neutral fermion
coupling to bottom-sbottom quarks, which is given in eq.~(\ref{F0bottomsbottomCoup}).
Specializing that coupling to the case when the neutral fermion is a neutrino one finds,
\begin{center}
\vspace{-50pt} \hfill \\
\begin{picture}(110,90)(0,23) 
\DashLine(10,25)(50,25){4}
\ArrowLine(50,25)(78,53)
\ArrowLine(78,-3)(50,25)
\Text(10,35)[]{$\tilde b_k$}
\Text(90,55)[]{$b$}
\Text(90,-5)[]{$\nu_j$}
\end{picture}
$
=\,i\,\Big[O^{b\nu\tilde b}_{Ljk}\frac{(1-\gamma_5)}{2}+
O^{b\nu\tilde b}_{Rjk}\frac{(1+\gamma_5)}{2}\Big],
$
\vspace{30pt} \hfill \\
\end{center}
\vspace{10pt}
with
\begin{eqnarray}
O^{b\nu\tilde b}_{Ljk} &=&
\eta_j \frac{2gt_W}{3\sqrt{2}} \xi_{j1} R^{\tilde b}_{k2} + \eta_j h_b \xi_{j3} R^{\tilde b}_{k1}
\equiv \eta_j \widetilde O^{b\nu\tilde b}_{Lk} \, \Lambda_j - \eta_j \frac{h_b R^{\tilde b}_{k1}}{\mu}
\epsilon_j\,,
\nonumber\\
O^{b\nu\tilde b}_{Rjk} &=& 
- \frac{g}{\sqrt{2}} \left( \xi_{j2} - \frac{1}{3} t_W \xi_{j1} \right) R^{\tilde b}_{k1}
+ h_b \xi_{j3} R^{\tilde b}_{k2}
\equiv \widetilde O^{b\nu\tilde b}_{Rk} \, \Lambda_j - \frac{h_b R^{\tilde b}_{k2}}{\mu}
\epsilon_j\,.
\end{eqnarray}
In the last equalities, the $\widetilde O$ couplings are defined as,
\begin{eqnarray}
\widetilde O^{b\nu\tilde b}_{Lk} &=& 
\frac{2gt_W}{3\sqrt{2}} \xi_1 R^{\tilde b}_{k2} + h_b \xi_3 R^{\tilde b}_{k1}\,,
\nonumber\\
\widetilde O^{b\nu\tilde b}_{Rk} &=& 
- \frac{g}{\sqrt{2}} \left( \xi_2 - \frac{1}{3} t_W \xi_1 \right) R^{\tilde b}_{k1}
+ h_b \xi_3 R^{\tilde b}_{k2}\,.
\end{eqnarray}
In addition, the neutralino coupling to bottom-sbottom, given in eq.~(\ref{X0bottomsbottomCoup}), 
is needed. The bottom-sbottom loops contributing to $\delta m$ are therefore,
\begin{center}
\vspace{-50pt} \hfill \\
\begin{picture}(110,90)(0,23) 
\ArrowLine(0,25)(30,25)
\ArrowArc(50,25)(20,180,0)
\ArrowLine(70,25)(100,25)
\DashCArc(50,25)(20,0,180){4}
\Text(51,53)[]{$\tilde b_k$}
\Text(50,-4)[]{$b$}
\Text(10,35)[]{$\nu_i$}
\Text(90,35)[]{$\chi^0_j$}
\end{picture}
$
=\,i\,\Sigma_{i+4,j}^{b\tilde b}(p^2),
$
\vspace{30pt} \hfill \\
\end{center}
\vspace{10pt}
with
\begin{equation}
\Sigma_{i+4,j}^{b\tilde b}(p^2) = -\frac{n_ch_b^2\xi_{i3}N_{j3}}{16\pi^2} \sum_{k=1}^2 \left[
m_b R^{\tilde b}_{k1} R^{\tilde b}_{k2} (\eta_iP_L+\eta_jP_R) B_0^{pb\tilde b} -
\slashed{p} (R^{\tilde b 2}_{k1}\eta_i\eta_j P_L + R^{\tilde b 2}_{k2} P_R) B_1^{pb\tilde b}
\right]+...
\end{equation}
where again only terms proportional to the Yukawa coupling squared are kept.
The above leads to the following contribution to $\delta m$,
\begin{eqnarray}
\delta m_{ij}^{b\tilde b} &=& \frac{n_ch_b^2\xi_3N_{j3}}{64\pi^2} \sum_{k=1}^2 \bigg\{
m_b s_{\tilde b} c_{\tilde b} (-1)^k \left( \eta_i+\eta_j \right) \Big[
B_0(m_{\chi_j^0}^2;m_b,m_{\tilde b_k}) + B_0(0;m_b,m_{\tilde b_k}) \Big]
\nonumber\\ && \qquad\quad
- m_{\chi_j^0} \left( \eta_i\eta_jc_{\tilde b}^2 + s_{\tilde b}^2 \right) 
B_1(m_{\chi_j^0}^2;m_b,m_{\tilde b_k})
\bigg\} \Lambda_i
\nonumber\\ &&
- \frac{n_ch_b^2N_{j3}}{64\pi^2\mu} \sum_{k=1}^2 \bigg\{
m_b s_{\tilde b} c_{\tilde b} (-1)^k \left( \eta_i+\eta_j \right) \Big[
B_0(m_{\chi_j^0}^2;m_b,m_{\tilde b_k}) + B_0(0;m_b,m_{\tilde b_k}) \Big]
\nonumber\\ && \qquad\quad
- m_{\chi_j^0} \left( \eta_i\eta_jc_{\tilde b}^2 + s_{\tilde b}^2 \right) 
B_1(m_{\chi_j^0}^2;m_b,m_{\tilde b_k})
\bigg\} \epsilon_i
\equiv (\delta m_{3,\Lambda}^{b\tilde b}) \Lambda_i + (\delta m_{3,\epsilon}^{b\tilde b}) 
\epsilon_i\,.
\label{dmBottomSbottom}
\end{eqnarray}
From this result one learns that the second order term in eq.~(\ref{OneLoopCorr}) can generate 
a solar neutrino mass from bottom-sbottom loops, because of the term proportional to $\epsilon_i$ 
in eq.~(\ref{dmBottomSbottom}). But that fact is known. More importantly, one learns that the top-stop
loops can contribute to the solar mass through the third order term in  
eq.~(\ref{OneLoopCorr}), in combination with the bottom-sbottom loops.

\section{Inverse Neutralino Mass Matrix}\label{AppB}

For the reader's benefit, the inverse of the tree-level neutralino mass matrix is given. Its 
matrix elements are equal to,
\begin{equation}
({\mathrm{M}}_{\chi}^0)^{-1} = \frac{1}{\det{M_{\chi^0}}} \left[ \begin{matrix}
I^{gg} & I^{gh} \cr I^{hg} & I^{hh}
\end{matrix} \right],
\end{equation}
with the following expressions for each sub-matrix,
\begin{eqnarray}
I^{gg} &=& \left[ \begin{matrix}
-M_2\mu^2+\frac{1}{2}g^2v_uv_d\mu & \frac{1}{2}gg'v_uv_d\mu \cr
\frac{1}{2}gg'v_uv_d\mu & -M_1\mu^2+\frac{1}{2}g'^2v_uv_d\mu
\end{matrix} \right],
\nonumber\\
I^{gh} &=& \left[ \begin{matrix}
-\frac{1}{2}g'v_uM_2\mu & \frac{1}{2}g'v_dM_2\mu \cr
\frac{1}{2}gv_uM_1\mu & -\frac{1}{2}gv_dM_1\mu
\end{matrix} \right],
\\
I^{hh} &=& \left[ \begin{matrix}
-\frac{1}{4}(g^2M_1+g'^2M_2)v_u^2 & M_1M_2\mu-\frac{1}{4}(g^2M_1+g'^2M_2)v_uv_d \cr
M_1M_2\mu-\frac{1}{4}(g^2M_1+g'^2M_2)v_uv_d & -\frac{1}{4}(g^2M_1+g'^2M_2)v_d^2
\end{matrix} \right],
\nonumber
\end{eqnarray}
and $I^{hg}=(I^{gh})^T$. 

\section{Approximated Neutralino/Neutrino Rotation Matrix}\label{AppC}

The neutralino/neutrino $7\times 7$ mass matrix is diagonalized, in first approximation, by
\begin{equation}
{\cal N}\approx\left[\begin{array}{cc}
N    & N\xi^T \\
-N_\nu \xi & N_\nu
\end{array}\right]
\label{approxN}
\end{equation}
where $N$ diagonalizes the $4\times 4$ neutralino sub-matrix, $N_\nu$ diagonalizes the $3\times 3$ 
neutrino sub-matrix, and the $3\times 4$ matrix $\xi$ is part of the block diagonalization 
described in eq.~(\ref{R0}).
The parameters $\xi_{ij}=(m\,{\mathrm{M}}_{\chi^0}^{-1})_{ij}$ are 
very important and have simple expressions,
\begin{eqnarray}
\xi_{i1} = \frac{g'M_2\mu}{2\det{M_{\chi^0}}} \Lambda_i\,, \,\qquad
&&\xi_{i2} = \frac{gM_1\mu}{2\det{M_{\chi^0}}} \Lambda_i\,, \,\qquad
\nonumber\\
\xi_{i3} = \frac{(g^2M_1+g'^2M_2)v_u}{4\det{M_{\chi^0}}} \Lambda_i-\frac{\epsilon_i}{\mu}\,, \,\qquad
&&\xi_{i4} =-\frac{(g^2M_1+g'^2M_2)v_d}{4\det{M_{\chi^0}}} \Lambda_i\,. \,\qquad
\end{eqnarray}
One defines also the reduce notation $\xi_{i1}=\xi_1\Lambda_i$, $\xi_{i2}=\xi_2\Lambda_i$,
$\xi_{i3}=\xi_3\Lambda_i-\epsilon_i/\mu$, and $\xi_{i4}=\xi_4\Lambda_i$.


\end{document}